\documentclass[conference]{IEEEtran}
\IEEEoverridecommandlockouts

\usepackage{cite}
\usepackage{ragged2e}
\usepackage{multirow}
\usepackage{wasysym}
\usepackage{tablefootnote}
\usepackage{graphicx}
\usepackage{xparse}
\usepackage{subfig}
\usepackage{comment}
\usepackage{booktabs}

\usepackage[hidelinks]{hyperref}

\usepackage{xcolor}

\usepackage{url}
\usepackage{xspace}
\usepackage[nolist]{acronym} 
\usepackage{balance}

\newcommand{\eg}{\textit{e.g.,~}}

\newcommand{\ie}{\textit{i.e.,~}}
\newcommand{\etal}{\textit{et al.}\xspace}
\NewDocumentCommand{\rot}{O{45} O{1em} m}{\makebox[#2][l]{\rotatebox{#1}{#3}}}

\newcommand{\delegcert}{\ac{DeCert}\xspace}
\newcommand{\delegcerts}{\acp{DeCert}\xspace}

\begin{document}

\title{Fine-grained CDN Delegation}

\author{\IEEEauthorblockN{Ethan Thompson}
\IEEEauthorblockA{
\textit{Carleton University}\\
Ottawa, Canada}
\and
\IEEEauthorblockN{Ali Sadeghi Jahromi}
\IEEEauthorblockA{
\textit{Carleton University}\\
Ottawa, Canada}
\and
\IEEEauthorblockN{AbdelRahman Abdou}
\IEEEauthorblockA{
\textit{Carleton University}\\
Ottawa, Canada}
}

\maketitle

\acrodefplural{RP}[RPs]{Relying Parties}

\begin{acronym}[itemsep=1ex]
    \acro{API}{Application Programming Interface}
    \acro{CA}{Certificate Authority}
    \acroplural{CA}[CAs]{Certificate Authorities}
    \acro{CDN}{Content Delivery Network}
    \acroplural{CDN}[CDNs]{Content Delivery Networks}
    \acro{CRL}{Certificate Revocation List}
    \acro{CSR}{Certificate Signing Request}
    \acro{CT}{Certificate Transparency}
    \acro{DANE}{DNS-Based Authentication of Named Entities}
    \acro{DC}{Delegated Credential}
    \acro{DDoS}{Distributed Denial of Service}
	\acro{DeCert}{Delegation Certificate}
    \acro{DNSSEC}{Domain Name System Security Extensions}
    \acro{DNS}{Domain Name System}
    \acro{DoNOR}{Distribution of Non-Owned Resources}
    \acro{DV}{Domain Validation}
    \acro{EEC}{End-Entity-Certificate}
    \acro{HTTPS}{Hypertext Transfer Protocol Secure}
    \acro{HTTP}{Hypertext Transfer Protocol}
    \acro{IANA}{Internet Assigned Numbers Authority}
    \acro{IdP}{Identity Provider}
    \acro{IETF}{Internet Engineering Task Force}
    \acro{IoT}{Internet of Things}
    \acro{IP}{Internet Protocol}
    \acro{OCSP}{Online Certificate Status Protocol}
    \acro{PitM}{Person-in-the-Middle}
    \acro{PKI}{Public Key Infrastructure}
    \acro{RP}{Relying Party}
    \acro{RFC}{Request for Comments}
    \acro{RP}{Relying Party}
    \acro{RR}{Recursive Resolver}
    \acro{SAML}{Security Assertion Markup Language}
    \acro{SAN}{Subject Alternative Names}
    \acro{SMS}{Short Messaging Service}
    \acro{SP}{Service Provider}
    \acro{SSH}{Secure Shell}
    \acro{SSL}{Secure Socket Layer}
    \acro{SSO}{Single Sign-On}
    \acro{STAR}{Short-Term Automatically Renewed}
    \acro{SUIT}{Software Updates for Internet of Things}
    \acro{SUP}{Software Update Provider}
    \acro{TLD}{Top Level Domain}
    \acro{TLS}{Transport Layer Security}
    \acro{TTL}{Time to Live}
    \acro{UI}{User Interface}
    \acro{URL}{Unifrom Resource Locator}
\end{acronym}

\begin{abstract}
The use of \acp{CDN} has significantly increased over the past decade, with approximately 55 million websites currently relying on CDN services.\footnote{\url{https://trends.builtwith.com/CDN/Content-Delivery-Network}
} Emerging solutions, such as Delegated Credentials (RFC 9345), lack fine-grained definitions of many critical aspects of delegation, such as the length of delegation chains, revocation mechanism, permitted operations, and a well-defined scope for said delegation. We present \delegcerts, which modify X.509 certificate standard and add new extensions to enable fine-grained CDN delegation. \delegcerts allow domain owners to specify delegated and non-delegated subdomains, and control the depth of delegation extended by \acp{CDN}, which provides flexibility in delegation management. But more importantly, \delegcerts are built on a new principle which provides full autonomy to domain owners--domain owners can issue \delegcerts fully independent of \acp{CA}, and thus have greater flexibility in policy control, including revocation methods. Such level of flexibility would be hard to match if \acp{CA} where to issue such certificates. Revoking a \delegcert revokes delegation. We discuss multiple revocation mechanisms for a \delegcerts balancing security, performance, and delegator control. We modify Firefox to support \delegcert (\ie proper validation) as a proof-of-concept, and test it to demonstrate the feasibility, compatibility of \delegcerts with browsers and TLS/HTTPS protocols. \delegcerts enhance the security, scalability, and manageability of \ac{CDN} delegation, offering a practical solution for Internet services.

\end{abstract}

\IEEEpeerreviewmaketitle

\section{Introduction}
\acp{CDN} started gaining popularity in the late 1990s to address the rapidly growing web traffic and server load, with Akamai pioneering content distribution for websites starting 1999~\cite{dilley_globally_2002, akamai}. As Internet usage grew, security threats escalated due to the absence of inherent security in most protocol designs~\cite{cheswick2003firewalls}. \ac{SSL} and \ac{TLS} protocols enabled secure content delivery over \ac{HTTPS}, supported by the web \ac{PKI}~\cite{allen_tls_1999},~\cite{rescorla_transport_2018}. The web PKI relies on trusted \acp{CA} to issue certificates binding domain names to cryptographic keys, ensuring encrypted and authenticated client-server connections. In 2018, Google Chrome’s update marked \ac{HTTP} connections as insecure, accelerating \ac{HTTPS} adoption~\cite{noauthor_secure_nodate_google}.

\ac{CA} certificates assure clients of a website’s authenticity and secure communication, using the keys bound to the domain name in the certificate~\cite{van2020computer}. However, enabling CDNs to serve content over HTTPS on behalf of domain owners introduces delegation challenges. In this case, the user connects to a server not managed by the domain owner, but the server still needs a certificate proving it represents the domain in order to use HTTPS. Liang~\etal~\cite{liang_when_2014} identified common practices used to enable \acp{CDN} to serve web content on behalf of a domain, such as private key sharing, which compromise security by granting \acp{CDN} excessive control~\cite{liang_when_2014}. Subsequent proposals, including Keyless TLS~\cite{noauthor_keyless_nodate, noauthor_keyless_2014} and Delegated Credentials~\cite{barnes_delegated_2023}, addressed some issues but fell short of their security goals or lacked fine-grained control~\cite{bhargavan_content_2017, chuat_sok_2020}.

Existing delegation mechanisms such as Delegated Credentials~\cite{barnes_delegated_2023} or other proposals~\cite{welch_internet_2004} often lack the property of giving the domain owner the ability to formally declare that their domain is being delegated to a CDN while remaining in control over said delegation. These mechanisms often over-authorize \acp{CDN}, allowing them to obtain valid certificates for entire domains or access sensitive subdomains and read encrypted user passwords~\cite{xin_quantifying_2023}. This lack of precise control undermines domain owners’ authority and security. 

We propose \delegcerts, a novel mechanism leveraging X.509 certificate extensions to enable fine-grained \ac{CDN} delegation. \delegcerts enable domain owners to effectively state limitations on the delegatee regarding delegation. DeCerts allow domain owners to specify authorized subdomains and limit delegation depth, eliminating CA dependency and enhancing transparency for users and browsers regarding the delegation of a domain. Efficient revocation is achieved through short-lived certificates, ensuring automatic expiration without additional infrastructure. Additional revocation mechanisms with more explicit control and overhead have also been introduced for \delegcerts.

We implemented a proof-of-concept in Firefox browser (Section~\ref{sec:Proof of Concept}) to demonstrate DeCerts’ feasibility, compatibility with TLS/HTTPS, and deployability by domain owners. We evaluate \delegcerts against Proxy Certificates~\cite{welch_internet_2004}, Delegated Credentials~\cite{barnes_delegated_2023}, and adapted approaches like X.509 Name Constraints~\cite{rfc5280}, using a framework of four security and four deployment properties (Section~\ref{sec:Technical Comparisons}). \delegcerts provide a secure, scalable solution for CDN delegation, addressing critical gaps in existing mechanisms. We believe that completing CDN delegation using \delegcerts opens up new unexplored avenues of delegating services and operations on the internet, such as subcontracting web services.

To summarize, based on the limitation of the previously proposed delegation schemes and the absence of a precise delegation scheme from domain owners to \acp{CDN}, this paper contributes a new delegation scheme, enabling fine-grained delegation from the domain owners to \acp{CDN} with flexible revocation techniques proposed for it. Additionally we develop a proof of concept implementation of the scheme demonstrating its modest changes required on the client (browser) side and feasibility and compatibility with the TLS/HTTPS protocols. Finally, we conduct a security and deployment analysis of \delegcerts and other similar CDN delegation techniques.

The rest of the paper is structured as follows. Section~\ref{sec:Background} provides background on \acp{CDN} and their underlying operation. Section~\ref{sec:Related Work} reviews related work on domain owner authorization techniques for \ac{CDN} content distribution and related attacks. Section~\ref{sec:Attacks and Design Goals} presents our threat model and design goals for secure delegation with \delegcerts. The details of \delegcerts are described in Section~\ref{sec:Delegation Certificates}, along with a discussion of changes required in practice to implement \delegcerts. We compare \delegcerts to Proxy Certificates, Delegated Credentials, and Name Constraints using a framework covering security and deployment properties in Section~\ref{sec:Technical Comparisons}. We provide further discussion on \delegcerts in Section~\ref{sec:Discussion}, where we also discuss additional delegation challenges. Finally, Section \ref{sec:Conclusion} concludes.
\section{Background}\label{sec:Background}
When visiting a website, many resources (\eg images, videos, and scripts) are often hosted not on servers controlled by the website owner but on distributed infrastructure managed by \acp{CDN}. Hosting all content directly would require the website owner\footnote{Throughout this paper, we use the terms \emph{website owner} and \emph{domain owner} interchangeably, as our focus in CDN delegation primarily concerns the distribution of web content.} to maintain servers capable of handling numerous concurrent connections and achieving near-continuous availability, which is costly and impractical for most websites. Instead, \acp{CDN} offer a scalable solution by hosting and delivering website resources on behalf of domain owners, ensuring high uptime and efficient content distribution.

\acp{CDN} are designed to offload the duty of serving web content from a single web server to a distributed system which is better designed to handle the intense amount of requests for content \cite{noauthor_what_nodate_cloudflare, noauthor_what_nodate_amazon, noauthor_what_nodate_akamai}. \ac{CDN} companies own a large number of servers distributed across many regions to decrease the latency in web requests and manage load balancing across domains. This distribution is done largely by the edge servers, which are the web servers which are at the ``edge'' of the network that clients will request content from. To obtain new content, edge servers contact cache servers operated by the \ac{CDN}. The cache servers store website data, including web pages and other content such as images and documents, from the domain owner's origin server (where they upload the content). For example, when a user, Bob, visits Alice’s website (\texttt{alice.com}), he is redirected to a CDN, Carol (\texttt{carol.com}), which delivers Alice’s content on her behalf. This process exemplifies delegation, where Alice authorizes Carol to serve content, highlighting the need for secure mechanisms to manage such delegation effectively.

\subsection{Routing}
There are three common ways Alice routes requests from Bob to Carol~\cite{liang_when_2014}: \ac{URL} rewriting, changing the \ac{DNS} CNAME record, and having Carol host Alice's domain (domain hosting).

\textbf{URL Rewriting.}  Alice modifies the URLs embedded in the content (\eg in HTML, CSS, or JavaScript) to redirect subsequent requests to the \ac{CDN}'s (\ie Carol's) edge servers instead of the origin. For example, instead of a link point to:

\centering
\texttt{https://alice.com/image.jpg}

\justifying
Alice would modify the link to correctly point to:

\centering
\texttt{https://alice.carol.com/image.jpg}.

\justifying
\textbf{CNAME.}
Unlike \ac{URL} rewriting, a domain owner configures a \ac{DNS} CNAME record to redirect traffic from their domain (\eg \texttt{alice.com}) to the CDN’s domain (\eg \texttt{alice.carol.com}) without altering the origin infrastructure. When a user requests content, the DNS resolves the alias to the CDN’s edge server, which delivers the cached content.

\justifying
\textbf{Domain Hosting.} Domain hosting by \acp{CDN} involves the \ac{CDN} acting as the authoritative name server for a domain owner’s \ac{DNS} records, enabling the \ac{CDN} to manage and serve content on behalf of the delegator. Instead of merely caching content, the \ac{CDN} hosts the domain’s DNS infrastructure, controlling \ac{DNS} responses. For example, a domain owner configures their domain (\eg \texttt{alice.com}) to use the CDN’s name servers (\eg \texttt{ns1.carol.com}). When users request content, the \ac{CDN}’s name servers resolve the domain to its edge servers that deliver cached or proxied content.

We focus on cases where the CNAME and Domain Hosting methods are used since \ac{URL} Rewriting requires that the website owner still maintains a web server to serve web pages for their domain which contain links pointing to a \ac{CDN}. Despite \ac{URL} Rewriting not requiring the same type of \ac{CDN} authentication step covered in the next section (not requiring the \ac{CDN} to possess a certificate valid for the domain), it lacks many of the previously discussed benefits that make \acp{CDN} desirable to website owners.

\subsection{Authentication}\label{sec:CDN Authentication}
The trouble with the above request routing mechanisms is when \texttt{https} is mixed in. In the case of CNAME and Domain Hosting (which is our focus), Bob expects to receive a certificate that is meant to prove that they are connected, securely, to Alice at \texttt{alice.com}. This means that if Carol is going to identify as \texttt{alice.com}, they need to be able to provide Bob with a valid certificate and be able to communicate securely with Bob using the information provided in the certificate. The three prevalent methods in practice to complete this is by using either a custom certificate, a shared certificate, or a \ac{DC}.

\textbf{Custom Certificate.} To enable a CDN to serve content over \ac{HTTPS}, a common practice, known as custom certificates, requires the domain owner, Alice, to obtain an X.509 certificate for \texttt{alice.com} from a \ac{CA} and share both the certificate and its private key with the CDN, Carol~\cite{liang_when_2014}. When a client, Bob, connects to Carol, she presents the \texttt{alice.com} certificate and uses the private key to establish a secure \ac{TLS} connection. However, this private key sharing risks key compromise, enabling potential misuse by the \ac{CDN}, motivating secure alternatives like \delegcerts.

\textbf{Shared Certificate.} Shared certificates enable a \ac{CDN} to serve content over \ac{HTTPS} for multiple domains using a single X.509 certificate issued by a \ac{CA}~\cite{liang_when_2014}. The \ac{CDN} lists authorized domains, such as \texttt{alice.com}, in the certificate’s \ac{SAN} extension. When a client, Bob, connects to the \ac{CDN}, Carol, she presents the certificate to authenticate as \texttt{alice.com} during the \ac{TLS} handshake. However, this approach grants the CDN broad authority over all listed domains, risking misuse if compromised and lacking fine-grained delegation control, motivating solutions like \delegcerts.

\textbf{Delegated Credentials.} Specified in RFC 9345 \cite{barnes_delegated_2023}, Delegated Credentials bind an identity to a public key using a newly defined document and not an X.509 certificate, enabling \acp{CDN} to serve content over \ac{HTTPS} without sharing the domain owner’s private key. \acp{DC} are short-lived (7 days by default) and are issued to \acp{CDN} by the domain owner, who signs the credential using a cryptographic signing key whose corresponding verification key is present in the \ac{CA}-issued X.509 certificate for the domain owner. The \ac{CDN} provides the credential to web browsers and is proof that the holder of the credential (the \ac{CDN}) is authorized to serve content on behalf of the domain that signed the credential. While Delegated Credentials enhance security by avoiding private key sharing, they lack fine-grained control over delegation scope, motivating solutions like \delegcerts.

An extension of delegated credentials is the concept of \emph{revocable Delegated Credentials}, introduced by Yoon~\etal~\cite{yoon2023delegation}. In this approach, the revocation status of delegated credentials is managed by the domain owner through DNS records, whose integrity and authenticity are ensured by DNSSEC. To revoke an issued delegated credential, the domain owner publishes its serial number as a subdomain entry in the DNS. This information can then be verified by the TLS client (\eg a web browser), thereby enabling revocation of delegated credentials.
\section{Related Work}\label{sec:Related Work}
In this section, we survey prior work on mechanisms for delegating content delivery to \acp{CDN}, focusing on their security and deployment properties. For these schemes, we highlight limitations that \delegcert addresses, such as lack of fine-grained control and flexible revocation. We cover related work on attacks on \acp{CDN} later in Section~\ref{subsec:Attacks} as it suits the context of our threat model in Section~\ref{sec:Attacks and Design Goals}.

\subsection{Standardized Mechanisms} 
Delegated Credentials (DCs), per RFC 9345~\cite{barnes_delegated_2023}, allow CDNs to serve content over HTTPS without private key sharing, using short-lived credentials signed by the domain owner. Delegated Credentials, presented during TLS handshakes, prove authorization for domains but lack precise delegation scope control, such as subdomain restrictions. Proxy certificates, defined in RFC~3820~\cite{welch_internet_2004}, originally designed to support delegation in distributed and grid computing environments by temporarily delegating credentials to other processes and nodes. Using this method in the context of \ac{CDN} delegation, the domain owner can issue a short-lived Proxy Certificate to the \ac{CDN} without having to share their private key or have a valid certificate with multiple unrelated domains on the \ac{SAN} list. 

Another certificate-based approach to secure delegation is by using \ac{STAR} certificates~\cite{sheffer_automatic_2021, sheffer_automatic_2021}. Initially specified in \ac{RFC} 8739 \cite{sheffer_support_2020}, \ac{STAR} certificates were proposed as an alternative solution to revoking certificates compromised private keys. The \ac{STAR} certificates are, as the name implies, short in validity period and stored in a dedicated server, which is not controlled by a delegated \ac{CDN}. Whenever the certificate needs to be provided to a client, it is requested from the dedicated server. Instead of going through a revocation process, which has its own challenges, a new certificate can simply be issued in place of a certificate with a compromised key, and the compromised certificate will be removed from the dedicated server. The usage of \ac{STAR} certificates for delegation is specified in \ac{RFC}~9115 \cite{sheffer_automatic_2021}. In the delegation usage, the delegated entity (the \ac{CDN}) creates a \ac{CSR} and sends it to the domain owner, who then forwards it to the \ac{CA}. While STAR reduces key exposure, it lacks fine-grained subdomain control and adds issuance overhead. 

\subsection{Deployed Practices}
Custom certificates require domain owners to share their X.509 certificate and private key with \acp{CDN}~\cite{liang_when_2014}, enabling HTTPS delivery but risking key compromise and misuse. Shared certificates list multiple domains in a single certificate’s \ac{SAN} field~\cite{liang_when_2014}, granting \acp{CDN} broad authority over all listed domains, vulnerable to misuse if compromised. Proposed by Cloudflare in 2014, Keyless SSL is a form of proxied \ac{TLS} which introduces a dedicated server to assist with performing \ac{SSL}/\ac{TLS} handshakes \cite{noauthor_keyless_nodate, noauthor_keyless_2014}. The dedicated server, known as the Key Server, holds the private key associated with the domain and performs the necessary decryption and signing needed for establishing the session key. In this way, Keyless SSL removes the need for \acp{CDN} to move the private key to the edge server to perform a handshake with a client. However, it introduces handshake delays and potential security flaws~\cite{bhargavan_content_2017}. Additionally, the Key Server may be owned and operated by the domain owner, the \ac{CDN}, or another third-party. Similar to Keyless SSL, proxied \ac{TLS} using LURK \cite{boureanu_lurk_2020} removes the need for a \ac{CDN} to possess the private key, and abstracts away parts of the TLS handshake which requires the private key to another entity (not the \ac{CDN}). While adding other security benefits, LURK adds an overhead to the \ac{TLS} handshake which is greater than Keyless SSL. Modifying Keyless SSL, 3(S)ACCE-K-SSL is a 3(S)ACCE secure variant of Keyless SSL\cite{bhargavan_content_2017}. While increasing the security of the protocol, 3(S)ACCE-K-SSL adds much overhead to Keyless SSL, increasing the needed steps to perform a \ac{TLS} handshake greater than that of LURK.

\subsection{Other Adaptable Schemes}
Mechanisms not designed for delegation can be adapted for \acp{CDN}. Utilizing \ac{DNSSEC}, \ac{DANE} binds the certificates of the domain to the \ac{DNS} records of the domain \cite{hoffman_dns-based_2012}. Specific details of the certificate, such as the serial number or hash, are stored in a \ac{DNS} TLSA record on the authoritative name server. When a client is provided a certificate for the domain, it can then confirm that the certificate is legitimate by performing the normal verification methods and then also by confirming the certificate's presence in the TLSA record. This enables domain owners to specify which certificates can be used for their domain by specifying them on a server that may be in their control, instead of on a public \ac{CT} log controlled by a \ac{CA} \cite{laurie_certificate_2021}. Detailed in \ac{RFC} 5280 \cite{rfc5280}, the \texttt{Name Constraints} extension constrains which domains a certificate can issue valid certificates for. These restrictions are defined using \texttt{permittedSubtrees} and \texttt{excludedSubtrees} fields in the extension. The \ac{RFC} specifies that the \texttt{Name Constraints} extension should only be present on a certificate that is issued to a \ac{CA} (a certificate that has the \ac{CA} flag set to \texttt{TRUE}), but this would not stop a \ac{CA} from issuing a \ac{CA} certificate with the \texttt{Name Constraints} extension to a domain owner. This would enable to domain owner to issue valid X.509 certificates for their domain and subdomains, and provide those certificates to delegated \acp{CDN}. Simple Public Key Infrastructure (SPKI) certificates~\cite{rfc2693}, designed for authorization, could theoretically be applied to support \ac{CDN} delegation but lack widespread adoption and \ac{TLS} compatibility.

Our coverage of related work focuses on delegation techniques which can be used to delegate a \ac{CDN} to distribute content for a website. There are other techniques that work towards enabling a middlebox to intercept a TLS connection with the mutual agreement of both ends, examples of these include \ac{SSL} splitting \cite{ssl-splitting}. For a full survey, see \cite{carnavalet_survey_2020}. The category of techniques where a middlebox exits in the middle of a \ac{TLS} connection between the website owner and a client may appear to be related to \ac{CDN} delegation, but it does not apply because there is not necessarily an active connection between a \ac{CDN} and the website owner. For most client requests, the \ac{CDN} will have a cache of the content available for the website.
\section{Threat Model and Design Goals}\label{sec:Attacks and Design Goals}
In this section, we propose four design goals for secure and fine-grained \ac{CDN} delegation, addressing limitations in existing mechanisms. We also present a threat model tailored to \delegcerts, capturing risks in domain owner-CDN interactions to ensure robust delegation.

\subsection{Design Goals for \acs*{CDN} Delegation}\label{sec:Design Goals}
There are two key processes involved in delegation:
\begin{itemize}
    \item \textbf{Claim}: Consists of the steps taken to setup the delegation between the Delegator and the Delegatee. The process may involve establishing parameters to be used in the communication process.
    \item \textbf{Proof}: How the relying party or Consumer of the delegation identifies the delegated party (Delegatee).
\end{itemize}

In some cases, the method by which the configuration process is conducted can make the communication process easier (or even complete itself). For example, URL Rewriting completes both the configuration and communication process at the same time.

We define four goals to be sought while designing a \ac{CDN} delegation method (below).

\textbf{G1. Issue Explicit Delegation.} The delegation method must allow the website owner to \textit{explicitly} delegate content distribution to a \ac{CDN}, ensuring authorized operation for specified domains.

\textbf{G2. Revoke Delegation of a CDN.} The delegation method must allow the website owner to void delegation in a timely manner to mitigate risks from compromised or untrusted \acp{CDN}.

\textbf{G3. Provide Fine-Grained Control.} The delegation method must allow website owners to impose limitations on the \ac{CDN}, and that these limitations are clearly communicated to the \ac{CDN}.

\textbf{G4. Delegation Transparency to Web Browsers.} The mechanism must communicate delegation details to web browsers via standard mechanisms, clearly specifying the delegator, delegatee, and scope.

\delegcerts will also provide the additional benefit of enabling website owners to allow a delegated \ac{CDN} to delegate other subordinate \acp{CDN}, and is described in Section \ref{subsec:Handling Extended Delegation}.

\subsection{Threat Model}
\ac{CDN} delegation involves three parties: the \textbf{Delegator}, \textbf{Delegatee}, and \textbf{Consumer}. We define them as follows:

\textbf{Delegator:} The ``source'' entity that wishes to delegate to another entity the duty of distributing its content on the source entity's behalf. This typically the website owner.

\textbf{Delegatee:} The ``destination'' entity that has been, or will be, delegated by a Delegator the distribution of web content on the Delegator's behalf. This is typically the \ac{CDN}.

\textbf{Consumer:} The ``end'' entity that looks to use the Delegator's service through the Delegatee, sometimes without being aware that it is communicating with a delegated entity. In this way, the Consumer is consuming the delegated service. This is typically a web-browser.

In the three-party delegation model described above, although they may be susceptible to misconfigurations and attacks, we assume the Delegator and the Consumer to be trusted entities. The Delegatee is considered to be either malicious or ``careless'' as it lacks inherent motivation to otherwise care about the content it distributes. Our focus is on the cases where the Delegator or Consumer is vulnerable to attacks made possible by insecure delegation, compromising security properties like authenticity and integrity.

More specifically to \acp{CDN}, an adversary may be motivated to exploit the insecure delegation to gain the ability to impersonate the delegated domain and then serve malicious content under their identity. In this case, we assume the adversary could be the \ac{CDN} itself, or leverage other vulnerabilities on the \ac{CDN} to gain unauthorized access to resources. Once the adversary has this access, they could control the content being delivered to users visiting a website that had delegated content distribution to the \ac{CDN}. They may then leverage this ability to serve malicious content under the compromised domain, inject advertisements for revenue, steal user data, or exploit the access in other ways.

Attacks on the domain validation stage of \ac{PKI} are out of scope, as \delegcerts (herein) and other delegation mechanisms (\eg Delegated Credentials, Proxy Certificates) rely on a trusted \ac{PKI}.  Once an adversary has a certificate for a website which they have obtained after either compromising a \ac{CA} or the domain validation process they would then be able to issue \delegcerts to either themselves, or any other \ac{CDN}. Such attacks require orthogonal \ac{PKI} solutions (\eg Domain Validation++~\cite{domain_validation} or domain validation using multiple vantage points~\cite{DVVantage}), while \delegcerts herein focuses on delegation-specific risks.

\subsection{Attacks}\label{subsec:Attacks}
\ac{CDN} delegation requires the delegator (website owner) to specify an ``Origin Server'' from which the delegatee (\ac{CDN}) fetches content that is not already cached. When a Consumer (browser) requests a resource, the \ac{CDN}’s edge server queries the origin server and caches the response for future requests. The question arises: \emph{how is this back-end request secured?} One might assume that the request follows the same procedure as a client-to-server \ac{HTTPS} communication, employing \ac{TLS} and certificate validation. However, this is not always the case. Table \ref{tab:CDN attack analysis} summarizes the missing design goals exploited by the attacks discussed below.

Shobiri \textit{et al.}~\cite{shobiri_cdns_2023} studied the failures of \ac{CDN} providers when verifying the identity of an origin server, which can lead to the distribution of resources on the domain owner's website, where such resources were not authorized for distribution. They investigated 14 \acp{CDN} and found that all of them were vulnerable to some form of \ac{PitM} attack due to the failure of the \ac{CDN} to verify the origin server. The three problem areas observed by the authors \cite{shobiri_cdns_2023} were that \acp{CDN} are not properly verifying the provided certificate, using/supporting weak security parameters, and the use of weak default options to new customers (including not verifying the origin server identity at all). We observe that the \ac{PitM} attacks were largely enabled by the Delegatee (the \ac{CDN}) failing to complete the delegated task securely (from a partial lack of design goal G3), and the Delegator's inability to enforce secure practices on the Delegatee.

Zhang \textit{et al.}~\cite{zhang_talking_2020} demonstrate in Talking with Familiar Strangers how the \ac{HTTPS} configurations of one server can impact another server given that those two servers are serving content for domains that share a certificate. Due to how the delegation is proven (a shared certificate), the Delegator relies on other Delegators concerning their security, since they cannot create specific security requirements for their domain and resources served on their behalf (lacking design goal G3). In this way, the Delegator may be unable to enforce specific policies for their domain.

Guo \textit{et al.}~\cite{guo_abusing_2018} study how \acp{CDN} fail to validate origin servers, discussing six ways this can be exploited by a malicious customer and demonstrate such cases. They focus on eight popular \acp{CDN} and find that all of them are vulnerable to some form of abuse due to failures in origin validation. When we consider our properties when analyzing the abuse cases outlined in the paper, we find that there are failures in design goals G1 and G4. Since the content being served by the \ac{CDN} was not delegated by the content owner (the configured origin server) there was a failure with the issuance of delegation (G1) as the owner did not wish to delegate to the \ac{CDN}. Since there is no mechanism for the client to know that the \ac{CDN} serving them content was not delegated to serve the content, there is a lack of design goal G4 (communication of delegation to the client).

\begin{table}[t]
    \caption{Attacks involving \acp{CDN} that either miss or have poorly implemented our design goals, along with an indication of which missed goals could have prevented each attack. An empty circle denotes that satisfying the design goal does not prevent the attack; a filled circle indicates that it would prevent the attack; and a half circle means satisfying the design goal may prevent the attack, or make it harder.}
    \resizebox{\linewidth}{!}{
    \begin{tabular}{l|c|c|c|c}
         \multirow{2}{*}{Attacks in the Literature} & \multicolumn{4}{c}{Enabled by the Lack of}\\
         & G1 & G2 & G3 & G4 \\
         \hline
         CDNs' Dark Side \cite{shobiri_cdns_2023} & \Circle & \Circle & \RIGHTcircle & \Circle \\
         Talking with Familiar Strangers \cite{zhang_talking_2020} & \Circle & \Circle & \CIRCLE & \Circle \\
         Abusing CDNs for Fun and Profit \cite{guo_abusing_2018} & \CIRCLE & \Circle & \Circle & \CIRCLE \\
         CDN Backfired \cite{li_cdn_2020} & \RIGHTcircle & \Circle & \CIRCLE & \Circle 
    \end{tabular}}
    \label{tab:CDN attack analysis}
\end{table}

Li \textit{et al.}~\cite{li_cdn_2020} demonstrate a novel \ac{HTTP} amplification attack based on poor \ac{DDoS} mitigation mechanisms and range request implementation vulnerabilities of \acp{CDN}. Since the configured origin server \textbf{may} or \textbf{may not} be owned by the entity configuring the \ac{CDN}, this attack could be launched against domains that did not delegate the \ac{CDN} to serve their content (thus, there \textit{could} be a lack of design goal G1). There is no mechanism for the \ac{CDN} to be told how to handle range requests, so there is a failure in design goal G3, and also the \ac{CDN} cannot be told to make changes to their behavior. However, it could be possible for the origin server to make changes to their response to specific range requests.

In Section \ref{sec:Technical Comparisons}, we compare our proposed solution to others with respect to the desired secure delegation goals (above), and find that the presented \delegcerts achieve all of our design goals, including one that is not met by other proposed solutions.
\section{Delegation Certificates}\label{sec:Delegation Certificates}
\delegcerts are based on X.509 certificates, containing all the fields of a regular X.509 plus a new extension that we propose herein, called \textit{Delegation Info} (Sec.~\ref{sec:Finer-Grained}). The delegated entity generates a public/private key pair, embeds the public key in a CSR, and sends it to the domain owner (the delegator). If all fields are acceptable to the owner's security policy, \eg Public Key Algorithm and Key length, Key Usage and Extended Key Usage, Certificate policies, Path Length Constraint, Subject Alternative Name (SAN), and upon validating the requester's possession of the corresponding private key, the owner issues the requester a \delegcert that binds the requester's public key to the owner's domain name. This allows the delegatee (the requester) to interact with a client on behalf of the domain owner, serving the \delegcert along with the rest of the domain owner's certificate chain to the client for validation.

As shown in Figure~\ref{fig:Delegcerts1}, suppose \texttt{abc.com} has an agreement with \texttt{cdn.com} to serve the subdomain \texttt{*.content.abc.com}. The CDN first generates a public/private key pair and sends the public key in a CSR to \texttt{abc.com}. If acceptable to \texttt{abc.com}, it issues a \delegcert that binds \texttt{cdn.com}'s public key to \texttt{*.content.abc.com}. The \delegcert has \texttt{cdn.com} as the Common Name (CN) under Subject Name, \texttt{*.content.abc.com} in the \textit{Delegation Info} as \texttt{Included} extension, and \texttt{abc.com} as the CN under Issuer Name. This authorizes \texttt{cdn.com} to terminate a TLS session with a browser whose URL bar shows \texttt{*.content.abc.com}, using \texttt{cdn.com}'s public/private key pair. The authorization is limited only to \texttt{*.content.abc.com}, and naturally expires with the certificate expiry. The issuer, \ie \texttt{abc.com}, may also include CRL end points in the \delegcert if it prefers to control revocation beyond natural expiry. \texttt{cdn.com} then presents the \delegcert, which is non-CA certificate, and the (parent) certificate issued to \texttt{abc.com}, which is also a non-CA certificate, along with the rest of the chain of CA certificates to a client visiting \texttt{*.content.abc.com}.

The above example demonstrated a simple delegation from a domain owner to a \ac{CDN} using \delegcerts. In the remainder of this section, we detail how other features of \delegcerts can be leveraged to achieve fine-grained delegation.

A summary of the main differences between a \delegcert and current PKI certificate practices follows.

\begin{itemize}
\item \textsl{Certificate Issuance.} \delegcerts are issued by the domain owner, rather than a CA. The \texttt{CA} flag/field under the standard X.509 \textit{Basic Constraints} extension is set to \texttt{False} in a \delegcert, despite not being a leaf certificate (\ie not an end-of-chain certificate).

\item \textsl{Certificate Body.} \delegcerts contain a new \textit{Delegation Info} extension. The rest of the X.509 standard fields can exist in a \delegcert, which is at the core of their flexibility compared to DCs.

\item \textsl{Certificate Validation.} In addition to standard certificate validation, a client (\eg browser) receiving a certificate chain that contains one or more \delegcerts must also validate information in the \textit{Delegation Info} extension. More on validation can be found in Sections~\ref{sec:Finer-Grained} and \ref{subsec:Handling Extended Delegation}.

\item \textsl{Certificate Expiry and Renewal.} The domain owner issuing a \delegcert is responsible for renewing it, not a CA. In contrast to CAs, the number of \delegcerts a domain owner is expected to issue is significantly smaller. \delegcerts can thus be short-lived, as they can be auto renewed using authenticated, low overhead renewal requests made over the Internet by the delegated entities to the domain owner (the issuer). 

\begin{figure}[t]
    \centering
    \includegraphics[scale=0.7]{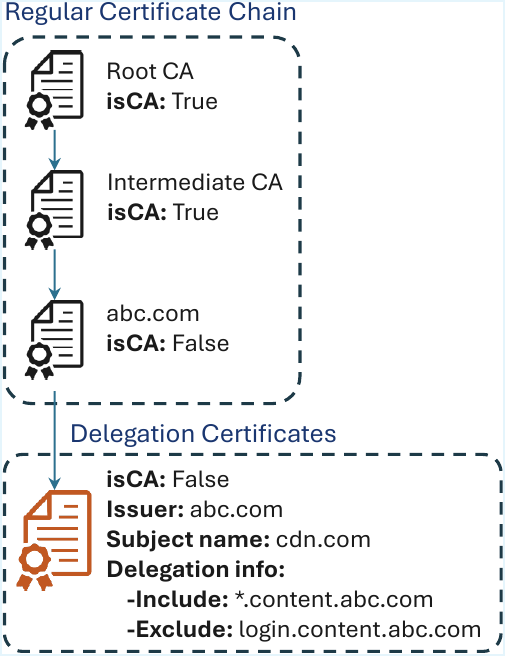}
    \caption{\delegcerts as an extension to the regular certificate chain, enabling fine-grained and flexible delegation to \acp{CDN} for content distribution.}
    \label{fig:Delegcerts1}
\end{figure}

\item \textsl{Path Length Constraint.} Under X.509 Basic Constraints, the \texttt{pathLenConstraint} applies only to CA certs, and is \textit{``the maximum number of non-self-issued intermediate certificates that may follow this certificate in a valid certification path.''}~\cite{rfc5280}. In our proposal, a \delegcert (which is a non-CA cert) can have a \texttt{pathLenConstraint} field, which would control the number of further delegations allowed. For example, if the domain owner prohibits the delegated entity from further delegating to others, it sets \texttt{pathLenConstraint=0}. More is in Sec.~\ref{subsec:Handling Extended Delegation}.

\item \textsl{Name Constraints} Under X.509 Basic Constraints, the Name Constraints extension applies only to CA certs. This field is not required in a \delegcert.

\item \textsl{Certificate Revocation.} The domain owner is responsible for revoking a delegation certificate, not a CA. While existing revocation techniques may be used (\eg OCSP stapling~\cite{rfc2560}), natural expiry of an \emph{ultra-short-lived} (\eg a few hours) certificate would be the most efficient way of revocation. While it can result in frequent renewals, it is not expected to overwhelm the domain owners because only delegated entities will be requesting renewals. Further discussion is in Sec.~\ref{ref:revocation}.

\item \textsl{CT logs and public search.} Unlike regular certificates, \delegcerts can only be issued by the domain owner (\ie other entities issuing them will lead to failure in client-side validation). As such, they need not be included in CT logs, thus preserving privacy of critical subdomains.

\end{itemize}

\subsection{Delegation Scoping}\label{sec:Finer-Grained}

To ensure that a domain owner can only issue certificates (\ie delegate) for her subdomains, the usage of the \textit{Delegation Info} extension requires analogous validation logic as the X.509 \texttt{Name Constraints} field~\cite{rfc5280}. This also applies to delegation chains (\eg a \ac{CDN} creating more sub-delegations), where a \delegcert issued by another \delegcert is scoped to only domains which the parent \delegcert is valid for. The parent \delegcert contains information that governs how long the \delegcert chain can be and is controlled by the issuing certificate belonging to the website owner.

\delegcerts contain a new extension \textit{Delegation Info} to enable the delegating domain owner to specify subsets of their domain to be included and excluded from the delegation. This field provides a finer scope to specify domains the certificate can be used for, without having to issue many certificates--one for each subdomain. Domain owners can clearly define the set of domains that are to be delegated by first specifying a large set of subdomains to include in the \texttt{Include} field of the Delegation Info, and then using the \texttt{Exclude} field to narrow down the scope and remove either specific subdomains from the set or a whole subset to be excluded from the larger include set. Therefore, the \texttt{Exclude} set of domains are applied after the \texttt{Include} set to restrict the set of delegated domains.

\begin{figure}%
    \centering
    \subfloat[\centering Invalid \texttt{PathLen} and \texttt{Include} \label{fig:invalid1}]{{\includegraphics[width=0.44\columnwidth]{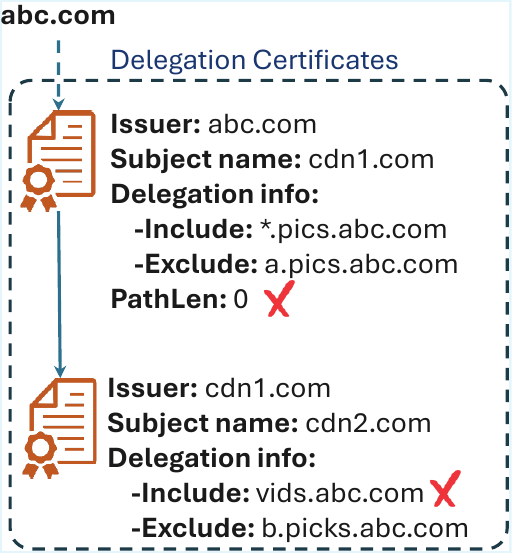} }}%
    \qquad
    \subfloat[\centering Invalid \texttt{Key usage}, \texttt{PathLen} and \texttt{Exclude} \label{fig:invalid2}]{{\includegraphics[width=0.44\columnwidth]{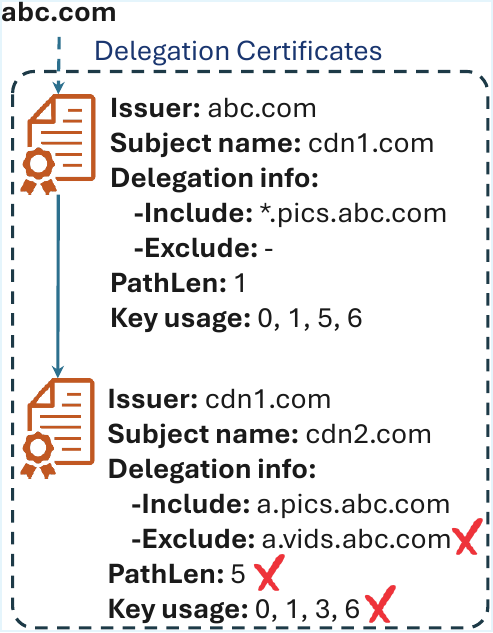} }}%
    \caption{\delegcerts and common forms of invalid delegation}%
    \label{fig:invalids}%
\end{figure}

As illustrated in Figure~\ref{fig:invalid1}, \texttt{abc.com} delegates all subdomains \texttt{*.pics.abc.com} except \texttt{a.pics.abc.com}; for instance, \texttt{figures.pics.abc.com} is delegated, whereas \texttt{a.pics.abc.com} is not. Note that to the best of our knowledge, all current methods (including Custom Certificates, Shared Certificates, and \acp{DC}) would require the website owner to authorize certificates specifically for each of the subdomains they use, causing management challenges and lack of flexibility with adding/removing new subdomains.

\subsection{Delegation Chains}
\label{subsec:Handling Extended Delegation}

Figure \ref{fig:invalid1} also illustrates how \delegcerts implement delegation chains, enabling a \ac{CDN} to partner with, \ie further delegate content distribution to, another \ac{CDN}. A new \delegcert is issued when further delegation is needed, creating a chain from the leaf certificate to the domain owner \delegcert. For example, as Figure~\ref{fig:invalid1} shows, suppose \texttt{abc.com} delegates \texttt{cdn1.com} to serve all of its subdomains for \texttt{pics.abc.com}. If \texttt{cdn1.com} subsequently delegates \texttt{cdn2.com} to serve only \texttt{log.pics.abc.com}, then \texttt{cdn1.com} issues a \delegcert to \texttt{cdn2.com} that binds \texttt{cdn2.com}’s public key to \texttt{logo.abc.com}. It is important to note that no subdomain beyond those originally delegated by \texttt{abc.com} can be further delegated; in this case, all subdomains are included except \texttt{a.pics.abc.com}.

Furthermore, the domain \texttt{abc.com} must specify a \texttt{pathLenConstraint} value of $1$ for the second DeCert to remain valid. In this example, however, the path length constraint is set to $0$; therefore, no further delegation is permitted, rendering the delegation depicted in Figure~\ref{fig:invalid1} invalid. In addition, the \texttt{Include} field in the first delegation authorizes only the subdomains of \texttt{pics.abc.com}, whereas the second delegation authorizes the subdomain \texttt{vids.abc.com}. Consequently, the second delegation is invalid, as \texttt{vids.abc.com} is not a subdomain of \texttt{pics.abc.com}.

In addition to the \texttt{Include} and \texttt{pathLenConstraint} fields, the \texttt{Exclude} and \texttt{KeyUsage} fields must also remain within the scope defined by their delegator’s DeCert. For instance, as illustrated in Figure~\ref{fig:invalid2}, the key usage specified in the delegation to \texttt{cdn2.com} is \{1, 3, 5, 6\}, where the value $3$ is not a subset of the delegator’s \texttt{KeyUsage} field. Furthermore, the \texttt{pathLenConstraint} of the second \delegcert is set to $5$, which is invalid since it exceeds the delegator’s \texttt{pathLenConstraint} value of $1$. Additionally, the \texttt{Exclude} field in the delegation information corresponds to \texttt{a.vids.abc.com}, which is not a subset of the delegator’s \texttt{Include} field that encompasses all subdomains of \texttt{pics.abc.com}. Therefore, the delegation depicted in Figure~\ref{fig:invalid2} is invalid for three distinct reasons.

Using the path length constraint field, the domain owner can specify how many certificates can be in the delegation chain. This allows the domain owner to control how far the delegation can extend. Each \delegcert in a chain is constrained/scoped by the issuing certificate. That is, each child certificate has no more capabilities (is valid for no more domains) than the parent certificate that issued it.

\subsection{Revoking a Delegation}
\label{ref:revocation}

Several revocation mechanisms can be employed for \delegcerts. The simplest approach involves issuing short-lived delegations, similar to those used in Delegated Credentials and \ac{STAR} certificates. In this method, \delegcerts are assigned a brief expiry window (\eg hours to days) after which they automatically and implicitly expire. While this is accompanied by the cascading invalidity problem (see Section \ref{sec:CascadingInvalidityProblem}), it requires no additional infrastructure changes, making implementation and adoption highly feasible. This eliminates the need for additional revocation inquiries during usage of \delegcerts, as the natural expiration handles revocation without requiring ongoing maintenance or status checks. 

For delegators requiring more explicit control, domain owners can implement traditional revocation mechanisms such as \ac{OCSP}, \acp{CRL}, or DNS-based revocation status publishing. These solutions enable domain owners to manage and publish the revocation status of their \delegcerts. However, they impact performance as these solutions introduce an additional round trip during the \delegcerts usage to query the revocation status. In the case of \acp{CRL}, the issue of large revocation lists is mitigated, as each domain owner maintains only a small \ac{CRL} specific to its limited number of \delegcerts. Similarly, if \ac{OCSP} is adopted, domain owners would operate their own \ac{OCSP} servers; while these servers would not face the overload experienced by CA-managed ones due to the constrained scope per domain, their high availability remains critical to prevent disruptions in certificate validation. Moreover, to improve performance and privacy, \ac{OCSP} stapling can be employed, allowing the revocation status to be sent alongside \delegcerts during TLS connections, avoiding the additional round trip caused by separate queries.

An alternative, relatively straightforward revocation solution is the DNS-based approach proposed by Yoon~\etal~\cite{yoon2023delegation} for Delegated Credentials, which leverages existing DNS infrastructure to publish revocation status by adding the serial numbers of revoked certificates as DNS records. This method is practical and fully manageable by domain owners, requiring only one additional DNS query during the TLS handshake to check the revocation status of \delegcerts. Overall, these mechanisms provide flexible options for revocation in \delegcerts, balancing simplicity, performance, and control while aligning with the decentralized nature of domain-owner delegations to \acp{CDN}.

\subsection{Changes Needed for Practical Adoption}
\delegcerts primarily benefit domain owners by providing fine-grained delegation with flexible revocation options (Section~\ref{ref:revocation}), without requiring changes to established web \ac{PKI} entities. This minimizes deployability challenges compared to CA-dependent mechanisms (\eg Proxy Certificates). Therefore, it is expected that the entities who initially push the adoption of the \delegcerts are the domain owners.

\acp{CA} are very well-established entities in the web \ac{PKI}. As such, any proposal that requires \acp{CA} to adopt some change can pose deployability challenges. \delegcerts do not require changes to be made to current leaf certificates or \acp{CA}, making their implementation that much more feasible for \acp{CDN}. To utilize \delegcerts, a \ac{CDN} would need to modify its automatic certificate management tool to renew \delegcerts when necessary.

The bulk of the changes needed for adoption are at the client browser level. Specifically, browsers would need to be modified to accept and validate \delegcerts. A non-supporting browser would show an error for an unknown extension on the certificate being marked critical. Once modified, the browser should be able to recognize the extension and be able to validate that the new extension is of the correct format (contains all required fields), is scoped within the issuing certificate, and meets all other standard certificate validation checks. See Section \ref{sec:Proof of Concept} for details on our implementation.

Other than the above modifications to the validation logic in the browser, no modifications are needed in the \ac{TLS} handshake. We consider this an advantage, as requiring a change to \ac{TLS} could introduce further complications to the adoption of \delegcerts.

Lastly, since \delegcerts can be implemented as an X.509 v3 extension, no modifications are needed to the current X.509 standard itself. The format of our X.509 v3 extension would need to be supported by the standards, as it is the defining trait of a \delegcert.

\subsection{Proof of Concept}\label{sec:Proof of Concept}
As a demonstration of our \delegcerts in action, we modified the Firefox Nightly version ``121.0a1'' browser to enable their usage. 17 files were modified, totaling 114 new lines of code. The vast majority of changes were made in two files, namely \emph{pkixnames.cpp} and \emph{pkixbuild.cpp}. We wrote scripts to create custom root and intermediate \ac{CA} certificates, as well as custom domain and \delegcerts. The custom root and intermediate \ac{CA} certificates were added to the root store~\cite{purushothaman2022position} of our modified browser. We setup a simple \ac{HTTPS} server locally to serve a simple web page, and connect to the domain using our modified browser.

We first test the behavior of the default, unmodified, Firefox browser by attempting to connect to our server using our \delegcert. In \delegcerts, the extension that defines a \delegcert is marked critical, as it is crucial that if the certificate is going to be used that a client receiving the certificate knows how to parse and validate the extension. In the case of an unknown extension marked critical, browsers are meant to not accept the certificates, as indicated in \ac{RFC} 5280 \cite{rfc5280}. As expected, the browser does not accept our certificate, throwing the SEC\_ERROR\_UNKNOWN\_CRITICAL\_EXTENSION indicating the presence of an extension on the certificate marked critical. In this case, the browser does not give the user the option to accept the certificate and connect to the website.

Next, we test the behavior of the Firefox browser after we have modified it to validate and accept \delegcerts which are issued by valid X.509 certificates, validating the entire chain from the \delegcert issue as normal. When connecting to our website, the browser receives, validates, and accepts our \delegcert and shows no additional messages in the browser \ac{UI} to the user, which is the same behavior as a regular website with a valid certificate.

To test Goal 3, fine-grained control (Section \ref{sec:Design Goals}), we issue another \delegcert which contains ``\url{*.a.localhost}'' in the \ac{SAN} extension and ``\url{b.a.localhost}'' in the \textit{excludeDomain} section of our \delegcert extension. This indicates that all subdomains of \url{a.localhost} have been delegated other than ``\url{b.a.localhost}'' and any of \url{*.b.a.localhost}. First, we attempt to connect to a valid domain using this certificate (\url{a.a.localhost}). Similar to our previous test, the domain receives, validates, and accepts the certificate correctly and allows the connection to the website without any additional action by, or indication to, the user. Analogous to normal TLS, the banner after the page load has been completed, does not indicate any issue with the certificate or connection. Next, we test that our modified browser will correctly warn about connections to a website using an invalid certificate. To do this, we attempt to connect to ``\url{b.a.localhost}'' (the domain indicated in the \textit{excludeDomain}). After receiving the certificate, the browser starts the validation process and finds that the certificate is not valid for the requested domain, as indicated in the \textit{excludeDomain} field of our extension. The browser then throws a warning to the user indicating that the certificate is not valid for this domain. The user then has the option to not connect to the website or to accept the certificate and continue. It is important to note that this is the same behavior that the browser shows when connecting to a website with other invalid X.509 certificates, which can be seen on \url{https://badssl.com}.

The proof-of-concept evaluation presented herein demonstrates the feasibility of \delegcerts and their compatibility with web browsers following minor modifications, without necessitating any alterations to the TLS protocol or the existing CA infrastructure.

\section{Comparative Evaluation}\label{sec:Technical Comparisons}
We compare \delegcerts to alternative methods that can be used for \ac{CDN} delegation: Delegated Credentials \cite{barnes_delegated_2023}, Proxy Certificates \cite{welch_internet_2004}, and Name Constraints \cite{rfc5280, chuat_sok_2020}. It is important to note that while Delegated Credentials are gaining increased industry attention, and now supported by CloudFlare~\cite{cloudflareDC}, Facebook~\cite{facebookDC}, and Firefox~\cite{firefoxDC}, the most implemented methods likely remain custom certificates (private key sharing) and shared certificates. Additionally, while Proxy Certificates are not adopted in the current WebPKI to the best of our knowledge, they were designed to provide many delegation benefits, albeit in a different environment (grid computing). We, therefore, choose to compare their benefits to \delegcerts.

\subsection{Meeting Design Goals}

\textbf{Goal 1.} As mentioned in Section \ref{sec:Design Goals}, our first goal is to provide website owners with the ability to delegate all or portions of their domain to a \ac{CDN}. For \delegcerts, this means simply issuing a new \delegcert to the delegated \ac{CDN}. The \ac{CDN} is named in the subject field of the certificate, and the delegated domains are captured in the \ac{SAN} and Delegation Info extension.

\textbf{Goal 2.} The revocation of a delegated \ac{CDN} relies on the ability of \delegcerts to be extremely short-lived, satisfying our second design goal. Specifically, a \ac{CDN} is only delegated as long as the certificate they hold is valid, which in the case of \delegcerts will be a short amount of time. This approach is similar to that used by Delegated Credentials \cite{barnes_delegated_2023} and STAR certificates \cite{sheffer_support_2020}. A benefit to this short-lived approach is that no additional infrastructure needs to be introduced to accommodate revoking/expiring \delegcerts. 

\textbf{Goal 3.} The third design goal captures the ability of the website owner to restrict what the delegated \ac{CDN} can do. In \delegcerts, the limitations of the \ac{CDN} are communicated on the certificate issued by the website owner. The excluded domains field is used to specify which domains this certificate cannot be used for, allowing website owners to retain more control over specific portions of their domain such as login or mail servers. Because the website owner controls all fields of the issued \delegcert, they can also dictate which policies are to be used and adhered to by the \ac{CDN} using the X.509 v3 Certificate Policies extension \cite{rfc5280}.

\textbf{Goal 4.} Lastly, our fourth design goal is to enable transparency to web browsers regarding the use of a delegated \ac{CDN}. Because \delegcerts are implemented through the use of X.509 v3 extensions, this allows them to retain all the benefits and richness of standard X.509 v3 fields and extensions. Most importantly, \delegcerts utilize the \ac{SAN} extension to communicate the scope of the delegation to the \ac{CDN}. This means that \delegcerts can instead utilize the \texttt{Subject} field of an X.509 certificate to communicate information about who the \ac{CDN} being delegated is without having to add an additional field. In this way, \delegcerts can communicate to web browsers that they are connecting to a delegated \ac{CDN}.

\begin{table}[h]
    \centering
    \caption{Various delegation solutions and the design goals they satisfy.}
    \resizebox{0.9\linewidth}{!}{
    \begin{tabular}{l|c|c|c|c}
        Proposal & G1 & G2 & G3 & G4 \\
        \hline
        \delegcerts & \CIRCLE & \CIRCLE & \CIRCLE & \CIRCLE\\
        Proxy Certificates \cite{welch_internet_2004} & \CIRCLE & \CIRCLE & \CIRCLE & \Circle\\
        Delegated Credentials \cite{barnes_delegated_2023} & \CIRCLE & \CIRCLE & \Circle & \Circle\\
        Name Constraints \cite{chuat_sok_2020} & \CIRCLE & \CIRCLE & \CIRCLE & \Circle
    \end{tabular}}
    
    \label{tab:Goal Analysis}
\end{table}

\subsection{Evaluating Other Schemes Against Our Design Goals}
In the next section, we compare \delegcerts, Proxy Certificates, Delegated Credentials, and Name Constraints, but do not include the previously mentioned practices involving custom and shared certificates. The reason behind this is that we do not consider the custom or shared certificate practices to be a formal delegation framework. That is, they are not designed to formally declare delegation from a website owner to a \ac{CDN}. We consider them a ``\textit{hack}'' to enable website owners to authorize a \ac{CDN} to serve content on their behalf using \ac{HTTPS}, but it does not effectively communicate who is being delegated to do what, and by whom. A clear and formal delegation solution would be able to answer questions such as ``Who is the delegated entity?'' ``Who is the original owner/Delegator?'' ``What are the limits of the delegation/what is the delegated entity allowed to do?'' We acknowledge that the shared certificates approach is closer to being able to answer the question of who is being delegated, but it still lacks the ability for the website owner to control other aspects of the delegation or efficiently communicate limitations such as which subdomains the \ac{CDN} is and is not allowed to serve content for. Because of this, we do not include custom or shared certificates in our comparisons.

\subsection{Comparative Deployment Evaluation}
Deployment considerations are vital when proposing new standards in an already existing and living infrastructure such as the internet. Here we consider four deployment attributes of various delegation solutions and compare them to how our proposed \delegcerts perform: server-side-modifications, client-side-modifications, \ac{TLS}-modifications, and \ac{CA}-support. Our analysis is illustrated in Table \ref{tab:Comparisons}.

\subsubsection*{Server-Side-Modifications}
Server-Side-Modifications describe modifications that will need to be made to the \ac{CDN} servers to utilize a given delegation solution. \delegcerts, Proxy Certificates, Delegated Credentials, and Name Constraints all require some form of change to be made by the \ac{CDN} to be used. In these cases, the \ac{CDN} would need to automate requests for the certificate/credential to the website owner and know how to store and use the certificate/credential. In the case of \delegcerts, Proxy Certificates, and Name Constraints, this would be the same process as using and requesting a regular X.509 v3 certificate, by making the request to the website owner rather than a \ac{CA}. For Delegated Credentials, the \ac{CDN} would need to use a new protocol to request and use the credential, as it is not built as an extension in X.509 certificates.

\subsubsection*{Client-Side-Modifications}
In all observed delegation solutions, modifications would need to be made to the client to implement the proposed solution. Both \delegcerts and Proxy Certificates would require the client to validate the newly presented extension, as well as accept certificates that are not issued by a \ac{CA} (though a valid \ac{CA} will be required in the trust chain). For Name Constraints, if the website owner were to only use their \ac{CA} issued certificate (that has the \ac{CA}-flag \texttt{TRUE}) to issue leaf certificates, then no modifications to the client would be needed. This is because the website owner now possesses a certificate with the \ac{CA} flag set to \texttt{TRUE}, and restricted using the \texttt{Name Constraints} field, browsers would not accept this as a leaf certificate. Assuming website owners would still wish to use this certificate as a leaf certificate, modifications would need to be made on the client side to accept these certificates. Delegated Credentials are a new document, not conforming to existing X.509 certificate standards. As such, Delegated Credentials require clients to be modified to recognize and validate the new document.

\subsubsection*{TLS-Modifications}
Because they are implemented in regular X.509 v3 certificates, \delegcerts, Proxy Certificates, and Name Constraints require no modifications to the \ac{TLS} protocol. Conversely, according to the Delegated Credentials \ac{RFC} \cite{barnes_delegated_2023} clients that are willing to accept Delegated Credentials are required to indicate so in a \ac{TLS} extension as part of their \texttt{ClientHello}.

\subsubsection*{CA-Support}
Our final deployment consideration is that of requiring \ac{CA} support. \acp{CA} are deeply rooted in the web \ac{PKI}, meaning their influence is great when proposing new standards in the web \ac{PKI}. Specifically, \acp{CA} are large stakeholders in the web \ac{PKI} and the issuing of certificates, as this is the foundation of their businesses. Solutions that aim to mitigate \ac{CA} usage may be viewed as being bad for business, and thus a \ac{CA} may not opt to allow them. Fortunately, \delegcerts and Proxy Certificates avoid logistical showstoppers by \acp{CA}, as they do not require any form of support from \acp{CA}. Delegated Credentials and Name Constraints however require \ac{CA} support. Delegated Credentials require a new X.509 v3 extension to be present on the website owner's certificate to indicate that it will be used for delegation. Name Constraints ask even more of \acp{CA}, being that they issue the website owner a certificate with the \ac{CA} flag set as true. Though implementing these changes may be trivial, if a \ac{CA} could detect that a Delegated Credential or Name Constraint solution was going to be used, they may choose to not issue such a certificate. In this scenario, \acp{CA} would be issuing fewer certificates (since the website owner can issue certificates that would previously have been issued by the \ac{CA}) which would result in the \ac{CA} making less money. While our \delegcerts also reduce the number of certificates issued by a \ac{CA}, the \ac{CA} is not able to distinguish between an \ac{EEC} that will be used for issuing \delegcerts or not, so they are not able to discriminate between them.

\subsection{Comparative Security Evaluation}
In this section, we observe the following security properties in the chosen delegation schemes: efficient revocation, controlled delegation chains, fine-grain delegation control, X.509 field control.

\subsubsection*{Efficient Revocation}
The reason a certificate may be revoked can vary, being motivated by stolen private keys or by the website owner wanting to cease their relationship with a previously delegated \ac{CDN}. In the case of \delegcerts, Proxy Certificates, and Delegated Credentials, they all rely on being short-lived in nature. This means that if the certificate/credential needs to be revoked, no additional steps would be taken as the certificate/credential will naturally expire soon. For \delegcerts and Proxy Certificates, the \ac{CRL} infrastructure could also be used as these are full X.509 certificates. Because Delegated Credentials are not X.509 certificates, changes would need to be made to the current \ac{CRL} infrastructure to accommodate the new document. The Name Constraints solution would not rely on the leaf certificates being short-lived per se, but instead on the regular X.509 revocation infrastructure including revocation lists and the \ac{OCSP} stapling practices.

\subsubsection*{Controlled Delegation Chains}
Currently, there are no delegation practices for \acp{CDN} that allow for delegation extension through a delegation chain. In some cases, this is a desired property, such as when a \ac{CDN} may wish to use a subordinate \ac{CDN} to help serve content for specific regions based on performance or other needs. For instance, consider the case where a country has laws in place to limit certain internet traffic through its borders. If a website owner still wishes for their web content to be available in that region, local CDN servers would need to be used. If the website owner has already delegated and configured another \ac{CDN} to serve their content, they would rather avoid having to make many changes when laws change in other countries. Instead, the \ac{CDN} may opt to partner with a \ac{CDN} local to that country or region to serve the web content on their behalf, and abide by the respective country laws. Of the delegation solutions observed, only our \delegcerts and Proxy Certificates enable a delegated party to further issue delegation to another entity. What the \ac{CDN} can further delegate is restrained by the set of domains the \ac{CDN} has been delegated to serve. Further, the length of the delegation chain can remain in the control of the website owner, as it can be indicated on the certificate they issue to the \ac{CDN} in the beginning. Neither Delegated Credentials nor Name Constraints allow for extended delegation, as neither would create valid chains if used for issuing more credentials/certificates.

\subsubsection*{Fine-Grain Subdomain Delegation Control}
In some cases, a website owner may wish to delegate only a portion of their domain to a \ac{CDN}, opting to remain in total control of specific subdomains. Covered in more detail in Section \ref{sec:Finer-Grained}, consider the case where a website owner wants a \ac{CDN} to serve content for all of their subdomains but wants to manage everything for their login server (such as \texttt{login.example.com}) without any other entity being involved in forwarding requests. Only \delegcerts easily allow for a website owner to delegate a portion of their domain rather than all subdomains. This is done through the use of the \ac{SAN} extension to indicate the larger set of domains to delegate and is restricted using a domain exclusion field as part of the delegation info extension. Proxy Certificates, Delegated Credentials, and Name Constraints do not allow for efficient partial domain delegation. That is, in their cases the website owner is forced to either delegate all of their domain or issue new certificates/credentials for each subdomain they wish to delegate.

This property is an adaptation of design goal 3, fine-grained control over delegated \ac{CDN} operations (\ref{sec:Design Goals}). Failure to achieve this goal in this case means that the \ac{CDN} would be ``over-delegated'' and now able to serve content for, and identify as, subdomains that the website owner may not have wanted to enable, such as the domain's mail server.

\subsubsection*{X.509 Field Control}
The richness of X.509 certificates enables greater communication regarding trust and identity aspects of a website. Further, X.509 v3 certificates allow for even greater coverage of information by allowing additional extensions to be added to a certificate to include information such as alternative names for the subject (\ac{SAN}), certificate key usage, and certificate policies. For a website owner to truly have control over how and what they delegate, it is essential that they have control over the X.509 fields of a certificate being used in delegation. Due to its lack of X.509 fields, \delegcerts are the only method observed that does not satisfy this condition.

Table \ref{tab:Comparisons} summarizes the above discussion. We also analyze the various attributes of these solutions, defined below, as well as observe which of our delegation goals the attribute satisfies.

\textbf{Implemented As.} How the delegation scheme is implemented. They are simply X.509v3 extensions, with the exception of Delegated Credentials which are not X.509 certificates, but rather a new type of document (or ``credential'').

\textbf{Conforms to SAN Usage.} If the delegation scheme is an X.509 extension, does it allow the usage of the \ac{SAN} extension?

\textbf{Server-Side-Modifications Not Required.} If the delegation scheme does not require any modifications to how CDNs process and handle certificates or credentials relative to their current operational practices, the scheme satisfies this property.

\textbf{Client-Side-Modifications Not Required.} If the delegation scheme does not require any modifications to how clients process and handle certificates or credentials in order to accommodate the proposed solution, the scheme satisfies this property.

\textbf{TLS-Modifications Not Required.} If changes are not required to be made to \ac{TLS} to accommodate the proposed solution, the scheme fulfills this property. 

\textbf{CA-Support Not Required.} If the support of \acp{CA} is not required to fully implement the delegation scheme, the scheme satisfies this property.

\textbf{Efficient Revocation.} How the delegation scheme handles the revocation of delegation.

\textbf{Controlled Delegation Chains.} Does the delegation scheme allow for the delegated \ac{CDN} to further delegate and issue valid delegations to other entities? An example of this would be if \ac{CDN} A were to partner with another \ac{CDN} (\ac{CDN} B) to distribute content in regions in which CDN A does not own accessible servers. \ac{CDN} B would need a valid certificate for the domains and content it is serving, even though the website owner was not the one to issue this delegation. Specifically, we are interested in solutions that do not rely on key sharing to complete this task.

\textbf{Fine-Grain Subdomain Delegation Control.} Is the website owner able to specify efficiently which subdomains are delegated to the \ac{CDN}?

\textbf{X.509 Field Control.} Does the website owner have the ability to control additional fields of the X.509 certificate issued to the \ac{CDN}?

\begin{table*}[ht]
    \caption{Comparison of delegation solutions and their deployment and security attributes.}
    \begin{center}
        \resizebox{0.9\linewidth}{!}{
        \centering
        \begin{tabular}{r|l|c|c|c|c|c}
             && \rot[90]{\delegcerts (herein)} & \rot[90]{Proxy Certificates} & \rot[90]{Delegated Credentials} & \rot[90]{Name Constraints} & Delegation Goal(s)\\
             \hline
             \multirow{2}{*}{Features}
             &Implemented As & X.509 Ext.~ & X.509 Ext.~ & New Document~ & X.509 Ext.~ & N/A\\
             &Conforms to SAN Usage & \checked & & & \checked & G1, G3\\
             \hline
             \multirow{4}{*}{Deployability}
             &Server-Side-Modifications Not Required &  &  &  &  & N/A\\
             &Client-Side-Modifications Not Required &  &  &  &  & N/A\\
             &TLS-Modifications Not Required & \checked & \checked & & \checked & N/A\\
             &CA-Support Not Required & \checked & \checked &  &  & N/A\\
             \hline
             \multirow{4}{*}{Security}
             &Efficient Revocation & \checked & \checked & \checked & & G2\\
             &Controlled Delegation Chains & \checked & \checked & & & G1\\
             &Fine-Grain Subdomain Delegation Control & \checked & & & \checked & G1\\
             &X.509 Field Control & \checked & \checked & & \checked & G3\\
             \bottomrule
        \end{tabular}
        }
        \label{tab:Comparisons}
    \end{center}
\end{table*}
\section{Discussion}\label{sec:Discussion}
In this section, we will discuss a less technical comparison between our work, Delegated Credentials, and Proxy Certificates, as well as two problems that are related to \ac{CDN} delegation, but we consider to be out of our scope.

\subsection{Further Comparisons with Delegation Certificates}
Compared to Delegated Credentials there are a few notable differences to \delegcerts illustrated in Tables \ref{tab:Comparisons}. The most prominent difference between Delegated Credentials and \delegcerts is how they are implemented. Delegated Credentials are separate from regular X.509 certificates which are used across the web. Instead, Delegated Credentials are implemented as a new type of document containing only a few fields that bind the document to the asserted identity and the issuing certificate. On the other hand, the herein presented \delegcerts are implemented as full X.509v3 certificates containing an additional extension to enable strong delegation properties. In both cases, the document is signed and issued by a valid certificate which was issued to the website owner by a \ac{CA}. Compared to \delegcerts, Delegated Credentials also lack the ability to describe finer-grained constraints on the delegation. Specifically, \delegcerts enable the Delegator to specify a set of domains which the certificate is valid for as well as how many certificates can be in the delegation chain from the issued \delegcert.

\delegcerts are very similar to Proxy Certificates in how they are implemented. That is, both are implemented as X.509v3 extensions. The overarching difference between Proxy Certificates and our \delegcerts is their intended application. Proxy Certificates were not designed with the Internet in mind, but rather within smaller networks where the ability to authorize automated tasks to act on behalf of the user is desired. A very important implementation difference between Proxy Certificates and \delegcerts is the use of the \ac{SAN} extension. Specifically, the Proxy Certificate specification prohibits the use of the \ac{SAN} field in a Proxy Certificate. This becomes very troublesome when some browsers rely on the \ac{SAN} field being present when matching the certificate with the requested domain, which is what we found when investigating the certificate validation process in Firefox. Alternatively, our \delegcerts allow, and actually require, the use of the \ac{SAN} extension as part of our implementation. This difference is more in line with how certificates are being used on the modern web and allows for a less restrictive implementation of \delegcerts.

\subsection{On DoNOR Attacks}
We define \ac{DoNOR} attacks as attacks that leverage an adversary's ability to have content that they do not own, and are not authorized to distribute, distributed. The issues stem from a lack of authenticating the resources an entity is claiming to have authority over before obtaining them. Shobiri \textit{et al.} studied this issue in \ac{CDN}s, finding 14 of the 14 \ac{CDN}s they investigated had back-end \ac{TLS} vulnerabilities that modern browsers could prevent/detect \cite{shobiri_cdns_2023}. They also found that 168,795 websites in the Alexa top 1 million may be vulnerable to back-end \ac{PitM} attacks between the distribution server and the origin server. We consider \ac{DoNOR} attacks to be outside of our scope, as our proposed solution relies on using secure \ac{TLS} practices.

\subsection{On The Cascading Invalidity Problem}\label{sec:CascadingInvalidityProblem}
The cascading invalidity problem is where you have a chain of certificates, each with differing validity windows (which is a safe assumption given a certificate doesn't usually get issued at the exact same time as its issuer). If the goal is to have a short validity period as a solution to revocation, the period in which each subsequent certificate in the chain (moving from parent to child) is valid is smaller and smaller, causing the overall ``work'' to issue certificates to eventually be unbearable. If the certificates are being distributed to different entities, and the ``head'' of the chain is only valid for a few hours, then given enough levels of the chain and certificates need to be issued every hour. Compound this with the time it takes to request and issue a new certificate and then propagate it to servers, multiplied by the number of unique domains and certificates, and this process can quickly become very expensive computationally.

DeCerts are designed with a short expiration window. To mitigate the \textit{Cascading Invalidity Problem}, certificates higher in the delegation chain may reuse the same key within this limited validity period following the expiration of a delegated DeCert. Consequently, DeCerts lower in the delegation hierarchy that possess longer validity periods and are signed with the same key remain valid. By renewing the certificates in the upper layers of the delegation chain (without updating their associated keys in short intervals) the cascading invalidity issue is effectively addressed. Moreover, the short-expiry design provides flexibility, allowing upper-level DeCerts to discontinue key reuse in the event of key compromise, thereby mitigating the associated security risks.

\section{Conclusion}\label{sec:Conclusion}
In this paper, we examined the existing challenges associated with \ac{CDN} delegation and highlighted the need for a secure delegation mechanism for website owners. Our analysis indicates that none of the currently available solutions fully satisfy the set of design goals outlined in Section~\ref{sec:Design Goals}. To address this, we introduced \delegcerts, a mechanism defined through a custom X.509v3 extension. We developed a proof of concept to demonstrate the feasibility of \delegcerts and conducted a comparative analysis of their security and deployment properties against existing approaches, including Delegated Credentials, Proxy Certificates, and Name Constraints. Furthermore, we discussed the minimal modifications required in the current Internet infrastructure to enable \delegcerts, emphasizing that only limited changes would be necessary for \acp{CDN} and web browsers as implemented in our prototype. Lastly, we also cover two challenges related to delegation on the Internet, being the first, to our knowledge, to categorize them as \ac{DoNOR} attacks and the Cascading Invalidity Problem respectively. We propose \delegcerts as a solution and a foundation for advancing research on secure and precise Internet delegation.

\balance
\bibliographystyle{IEEEtran}
\bibliography{references}

\end{document}